# A Routine for Measuring Synergy in University-Industry-Government Relations: Mutual Information as a Triple-Helix and Quadruple-Helix Indicator

Loet Leydesdorff,*[a] Han Woo Park,[b] & Balazs Lengyel[c]

**Abstract**
Mutual information in three (or more) dimensions can be considered as a Triple-Helix indicator of synergy in university-industry-government relations. An open-source routine th4.exe makes the computation of this indicator interactively available at the Internet, and thus applicable to large sets of data. Th4.exe computes all probabilistic entropies and mutual information in two, three, and, if available in the data, four dimensions among, for example, classes such as geographical addresses (cities, regions), technological codes (e.g., OECD's NACE codes), and size categories; or, alternatively, among institutional addresses (academic, industrial, public sector) in document sets. The relations between the Triple-Helix indicator—as an indicator of synergy—and the Triple-Helix model that specifies the possibility of feedback by an overlay of communications, are also discussed.

---

[a] University of Amsterdam, Amsterdam School of Communication Research (ASCoR), Kloveniersburgwal 48, 1012 CX Amsterdam, The Netherlands; loet@leydesdorff.net ; http://www.leydesdorff.net ; * corresponding author.
[b] Department of Media & Communication, YeungNam University, 214-1, Dae-dong, Gyeongsan-si, Gyeongsangbuk-do, South Korea, Zip Code 712-749; hanpark@ynu.ac.kr.
[c] International Business School Budapest, Tárogató út 2-4, 1021 Budapest, Hungary; blengyel@ibs-b.hu .

**Introduction**

The Triple Helix model of university-industry-government relations (Etzkowitz & Leydesdorff, 1995 and 2000) rapidly became interesting to policy makers because one can also interpret this program of studies as a call for greater collaboration locally. However, both the sciences and markets develop globally. The task of government is to retain wealth from knowledge or knowledge from wealth (e.g., oil revenues) by creating institutional frameworks which enable agents (firms, scholars) in regions and nations to be both global and local players, but with potentially different roles (Freeman & Perez, 1988). Thus, a trade-off between international and national/regional collaboration and orientation is required which is more difficult than a simple recipe in either direction. Some national systems, for example, are insufficiently open to global developments, while others are perhaps not sufficiently integrated internally. Note that the use of an indicator always requires theoretical interpretation.

Case studies inform us about best practices and one can learn from failures, but the results remain at the phenomenological level; the "phenotypes" of the triple helix partners (university, industry, government) can be expected to carry genotypical functions such as novelty production (in science), wealth generation (on the market), and governance at different scales (Ivanova & Leydesdorff, 2012; cf. Hodgson & Knudsen, 2011). From the perspective of the Triple-Helix theory, an indicator should ideally enable us to distinguish between a vicious lock-in into existing densities of relations (Biggiero, 1998) and too loose—or "footloose" (Vernon, 1979)—relationships at the global level that are insufficiently embedded locally (Cohen & Levinthal, 1990). Furthermore, such an indicator should not predicate specific levels of integration (at the national or regional level), but make this question empirical. The indicator should allow for extension to more than three dimensions. In other words, "systemness" can be considered as a hypothesis that can be tested in terms of the data. Is more "systemness" developed locally, regionally, sectorially, or nationally; and in which dimensions? (Carlsson, 2006).

We wish to contribute to this special issue about "Mapping Triple Helix Innovation for Developing and Transitional Economies: Webometrics, Scientometrics, and Informetrics," by making a routine available as open-source at the Internet for the computation of mutual information in three (or four) dimensions. Mutual information in three (or more) dimensions can be positive or negative and thus indicate synergy or a lack thereof in Triple-Helix (or Quadruple Helix) data. Increased access to large data sets, in our opinion, makes it urgent to develop tools that can be used flexibly to analyze data in different dimensions and using different units of analysis (such as firm data, publication data, etc.; cf. Park, 2010). The Triple-Helix indicator emerged from discussions about Triple-Relations among universities, industries, and government agencies, but its use is in no way confined to such relations alone.

**The Triple Helix Indicator of synergy in university-industry-government relations**

Building on Ulanowicz' (1986, at p. 143) proposal to use mutual information in three dimensions as an indicator of next-order systemness among three (or more) independent dimensions in the data, Leydesdorff (2003) suggested using university, industrial, and public-sector addresses as provided by the *Science Citation Index*, and thus to distinguish among nations in terms of the synergy in these institutional relations. Synergy can then be considered as a reduction of uncertainty in university-industry-government relations.

Two of the three dimensions can be partially and/or spuriously correlated in the third dimension. Spurious correlation can reduce uncertainty without being visible in the data without analysis—and therefore latent; whereas partial correlation can be measured using statistics such as Pearson's $r_{xy|z}$, or other measures such as Shannon's $H_{xy|z}$ as a measure of uncertainty in two dimensions conditioned on a third, and the corresponding mutual information $T_{xy|z}$.

Mutual information in more than two dimensions can be derived from Shannon's (1948) formulas (e.g., Yeung, 2008, at pp. 59f.), but it can no longer be considered as a Shannon entropy (Krippendorff, 2009). Its value can be positive, negative or zero, whereas Shannon types of information are necessarily positive (Theil, 1972). It can be shown (e.g., Abramson, 1963, at pp. 128 ff.; McGill, 1954) that the mutual information in or transmission (*T*) among three dimensions ($T_{xyz}$) is equal to:

$$T_{XYZ} = H_X + H_Y + H_Z - H_{XY} - H_{XZ} - H_{YZ} + H_{XYZ} \tag{1}$$

in which formula $H_X = -\sum p_X \log(p_X)$ and $H_{XY} = -\sum p_{XY} \log(p_{XY})$, etc., for one or more discrete variables *X* and *Y*, etc. Using the two-base for the logarithm, all these uncertainty measures are expressed in bits of information.

The further extension to more than three dimensions is straightforward. For example, Leydesdorff & Sun (2009; cf. Kwon *et al.*, 2011) considered the mutual information in the four dimensions of university (*u*), industry (*i*), government (*g*), or foreign (*f*) co-authorship relations in the case of Japan, using:

$$\begin{aligned} T_{uigf} = &\ H_u + H_i + H_g + H_f - H_{ui} - H_{ug} - H_{uf} - H_{ig} - H_{if} - H_{gf} \\ &+ H_{uig} + H_{uif} + H_{ugf} + H_{igf} - H_{uigf} \end{aligned} \tag{2}$$

These scientometric studies (see also: Park *et al.*, 2005; Park & Leydesdorff, 2010; Ye *et al.*, 2012) considered publications as units of analysis. A series of econometric studies was also developed using firms as units of analysis and firm size, technological classification, and address information as proxies for the economic, technological, and governmental dimensions, respectively. These studies comprise: Leydesdorff *et al.* (2006) about the Netherlands; Leydesdorff & Fritsch (2006) about Germany; Lengyel & Leydesdorff (2011) about Hungary; Strand & Leydesdorff (in press) about Norway; Leydesdorff & Strand (in press) about Sweden; and Perevodchikov & Leydesdorff (in preparation) about the Russian Federation.

**Automating the Triple-Helix Indicator**

The first version of an automated routine that allows users online to compute values for the TH Indicator—that is, mutual information in three dimensions—was made in 2008 and is available at http://www.leydesdorff.net/th .

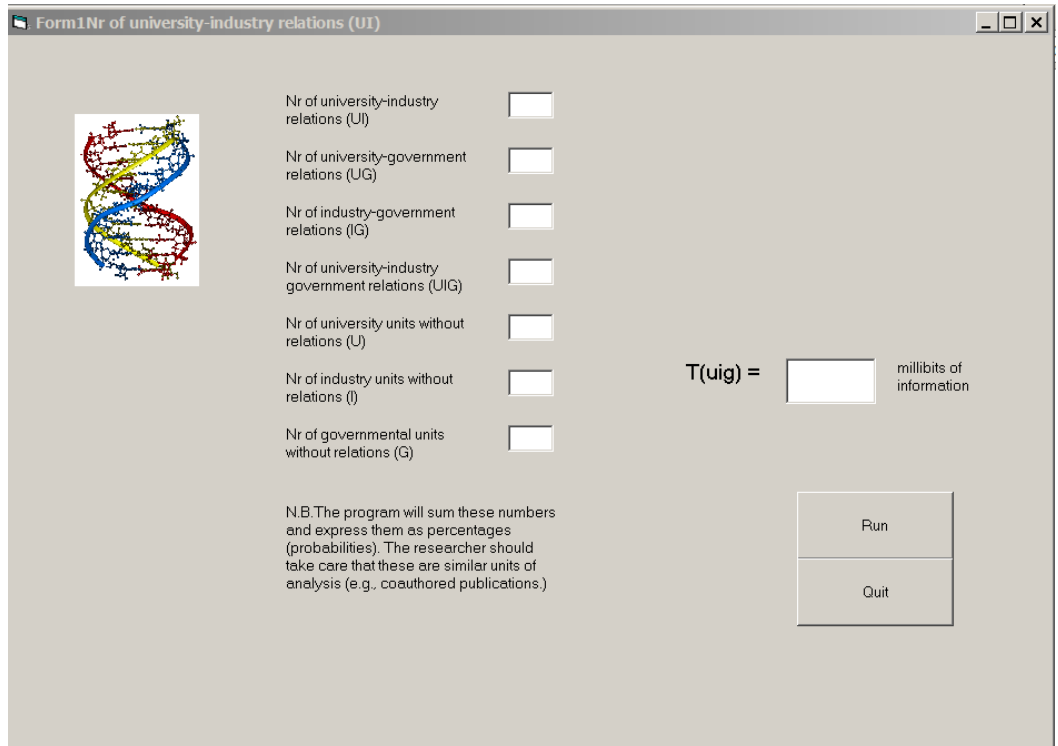

**Figure 1**: The routine TH.exe available at http://www.leydesdorff.net/th .

In 2009, this routine was elaborated with Krippendorff's (1980) $I_{ABC \to AB,AC,BC}$ as a measure for the uncertainty that is added by a three-dimensional interaction to the sum total of information contents of the two-dimensional interactions. Krippendorff's $I_{ABC \to AB,AC,BC}$, however, is a Shannon-type information that is necessarily positive (Leydesdorff, 2010 and 2011). One can also use Occam 3 for its computation (at http://occam.research.pdx.edu/occam/weboccam.cgi?action=search).

Following Krippendorff (2009b), the TH indicator—that is, mutual information in three dimensions—can be considered as the difference between the redundancy generated in the loops of communication possible in three-way interactions, and the (Shannon-type) information generated in these interactions. If the TH indicator is negative in terms of (e.g., bits of) information, this indicates a surplus of redundancy being generated. From the perspective of the TH model, this redundancy—that is, negative information or reduction of uncertainty—can also be considered as a measure of synergy at the systems level. In the TH model, this is considered as an "overlay" of communications which can feed back as a hyper-cycle on top of the cycles of communication within each of the helices. The hyper-cycle is a consequence of interactions among the relations between the underlying systems.

In other words, if universities and industries are already collaborating, the role of government can be different from a configuration where the two spheres operate at a distance. In a system with three functions, the order of operations can also be expected to make a difference: if government takes the lead, the result is different from leaving the lead to industry (Ivanova & Leydesdorff, in preparation).

The first routine (TH.exe) can be used for single values of the seven relevant variables: U, I, G, UI, UG, IG, and UIG. In most cases, these will be count data in document (or other) sets. The new routine th4.exe (at http://www.leydesdorff.net/software/th4) provides the user with the option to use batches of cases and up to four independent dimensions (Carayannis & Campbell, 2009, 2010; Leydesdorff, 2012). Each case (for example, a publication or firm) is represented as a single line. Each line first describes the case using an identifier (e.g., a sequence number), a value for potentially four variables called *w, x, y,* and *z*. If the fourth variable *z* is not specified, a single TH between *w*, *x*, and *y* is computed and all combinations with *z* are set equal to zero.

The variables and identifiers are considered as nominal variables. An input file has to be called "data.txt" and can, for example, be shaped as follows:

```
"id1", "1", "b", "region1", "2"
"id2", "2", "a", "region2", "1"
"id3", "1", "a", "region2", "2"
"id4", "1", "b", "region5", "1"
```

**Table 1**: Fictitious example of input data for th4.exe.

In the case of university-industry-government relations, for example, one may wish to code the second variable "Y" when a university address is present and "N" when not; and, *mutatis mutandis*, the third variable for industry addresses, etc. This file ("data.txt") should be stored in the same folder as th4.exe. The latter program generates a file th4.dbf in which the values for all relevant parameters are stored (Table 2). (The dbf-file can be read into such programs as Excel, SPSS, or OpenOffice.)

| Entropy and Transmission values in bits of information | |
| --- | --- |
| H(W) | 0.81 |
| H(X) | 1.00 |
| H(Y) | 1.50 |
| H(Z) | 1.00 |
| H(WX) | 1.50 |
| H(WY) | 2.00 |
| H(WZ) | 1.50 |
| H(XY) | 1.50 |
| H(XZ) | 2.00 |
| H(YZ) | 2.00 |
| H(WXY) | 2.00 |
| H(WXZ) | 2.00 |
| H(WYZ) | 2.00 |
| H(XYZ) | 2.00 |
| H(WXYZ) | 2.00 |
| T(WX) | 0.31 |
| T(WY) | 0.31 |
| T(WZ) | 0.31 |
| T(XY) | 1.00 |
| T(XZ) | 0.00 |
| T(YZ | 0.50 |
| T(WXY) | 0.31 |
| T(WXZ) | -0.19 |
| T(WYZ) | -0.19 |
| T(XYZ) | 0.00 |
| T(WXYZ) | -0.19 |

**Table 2**: Output of th4.exe on the basis of the data in Table 1, in bits of information.

For example, the second variable (that is, the first dimension *w*) in Table 1 has three times the value of "1", and once a "2". The uncertainty $H$(w) in this distribution is consequentially – (3/4) $\log_2$ (3/4) – (1/4) $\log_2$ (1/4) = 0.31 + 0.50 = 0.81 bits of information. Table 2 shows furthermore that two of the three possible mutual informations in three dimensions are negative in this case, and the four-dimensional *T*(wxyz) would also be -0.19 bits.

This above example is fictitious. Table 3 provides the data of six among 14,552 firms in the Trondheim (Tromso) region with their firm identity number, geographical code, NACE-code, and size code, as used by Strand & Leydesdorff (in press).

```
459695,1901,5,3
459696,1901,5,5
459697,1901,11,1
459698,1901,11,2
459699,1901,11,2
459700,1901,11,2
```

**Table 3**: Six firms in the Tromso region (coded "1901") with their identity-codes in the first column, and NACE-codes (OECD) and size codes in columns 3 and 4, respectively.

When one feeds the entire set of 14,552 firms into th4.exe, the resulting value of T(wxy) = -0.135 bits of information. Note that the fourth variable *z* was not specified in this set. As Table 3 indicates, the quotation marks are not strictly needed, but they are convenient in order to remember the nominal character of this data. In the future, we plan to make a similar program for numerical data, but in many applications the data at the case level will be coded into categories.

The output is collected in a file named "th4.dbf" that can be read into Excel, SPSS, or OpenOffice. If this file does not yet exist, it is generated by the routine; otherwise, a new record is added each time this routine is used. Thus, one can run subsets (e.g., regions) consecutively (by renaming files into "data.txt" in the right order), and each region will be represented as a row in the file th4.dbf.

**Conclusions and discussion**

The Triple-Helix indicator fulfills the requirements that were specified in the introduction on the basis of the Triple-Helix model: the trade-off between a relative rigidity which may emerge in an institutional network that is too densely populated with links, versus too much volatility in a network that is internationally oriented, can be quantified in terms of this *signed* information measure. Furthermore, there are virtually no size limitations for the data. (The current program is limited to 2 GByte of data so that it is compatible with both a 32-bits and 64-bits operating system.) However, the interpretation remains a bit elusive: a negative information is a redundancy. This redundancy is generated in next-order loops in the information processing which remain hypothesized, but cannot be observed directly (Luhmann, 1984, p. 226; 1995, p. 164).

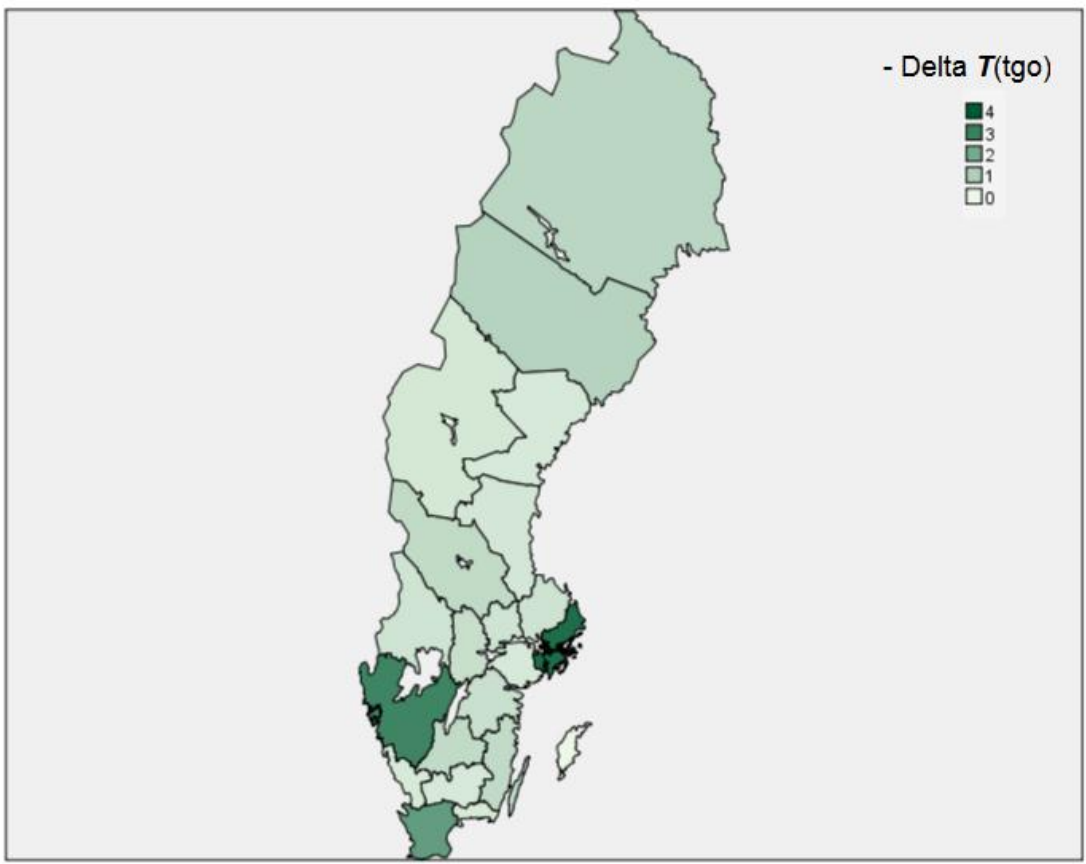

**Figure 2**: Contributions to the reduction of uncertainty at the level of 21 Swedish counties. (ΔT among technology, geography, and organization; Source: Leydesdorff & Strand, in press.).

In other words, one can hypothesize the knowledge base of an economy to be structured at the national, regional, or other levels, and then test these hypotheses in terms of the Triple-Helix indicator. Figure 2, for example, provides such a map for Sweden. The results can be mapped onto the geography after decomposition (contributions to the synergy are indicated as $-\Delta T_{TGO}$ in Figure 2), but one can also decompose in the other dimensions (Theil, 1972), for example, between high-tech manufacturing and knowledge-intensive services (e.g., Leydesdorff *et al*., 2006) or between national and international collaborations (Leydesdroff & Sun, 2009). The test-statistics in terms of significance are as yet poorly developed for entropy statistics. Although one is able to relate information theory to likelihood ratios between alternative hypotheses and Bayesian statistics (Sheskin, 2011, at pp. 384 ff.; Leydesdorff, 1992 and 1995), the major application of information theory remains oriented towards descriptive statistics of evolving systems (Leydesdorff, in press).

The further extension to more than three dimensions is popular among scholars who work with the Triple-Helix model (e.g., Bunders *et al*., 1999; Carayannis & Campbell, 2010 and 2011; Leydesdorff & Etzkowitz, 2003). Equations (1) and (2) show that the sign may change, when one extends from three to four (or more) dimensions. Krippendorff (2009b, at p. 670) formalized this in terms of the cardinality of a parameter $\Gamma$ in a more difficult summary equation. However, the interpretation of these next-order terms may become increasingly difficult (Leydesdorff, 2012).

Nevertheless, we hope to have provided a service to the community of Triple-Helix researchers by making these programs available on the internet and as open source. One can also use them to check one's own calculations, for example, in Excel. Feedback from the side of users will be appreciated and will hopefully lead to further improvements.

**Acknowledgement**
We acknowledge Øivind Strand for providing data about Tromso.